\documentclass{pasj00}

\begin{document}
\SetRunningHead{K.~Nagashima et al.}
{Hinode/SOT Observations of Sunspot Oscillations}
\Received{2007/06/12}
\Accepted{2007/09/03}

\title{Observations of Sunspot Oscillations
in G band and Ca \emissiontype{II} 
H line with Solar Optical Telescope on Hinode}


\author{%
Kaori \textsc{Nagashima}\altaffilmark{1},
Takashi \textsc{Sekii}\altaffilmark{1,2},
Alexander G. \textsc{Kosovichev}\altaffilmark{3},
Hiromoto \textsc{Shibahashi}\altaffilmark{4},
Saku \textsc{Tsuneta}\altaffilmark{2}, 
Kiyoshi \textsc{Ichimoto}\altaffilmark{2}, 
Yukio \textsc{Katsukawa}\altaffilmark{2},
Bruce W. \textsc{Lites}\altaffilmark{5}, 
Shin'ichi \textsc{Nagata}\altaffilmark{6},
Toshifumi \textsc{Shimizu}\altaffilmark{7}, 
Richard A. \textsc{Shine}\altaffilmark{8}, 
Yoshinori \textsc{Suematsu}\altaffilmark{2}, 
Theodore D. \textsc{Tarbell}\altaffilmark{8}, 
and
Alan M. \textsc{Title}\altaffilmark{8}}
\altaffiltext{1}{Department of Astronomical Science, 
The Graduate University for Advanced Studies (Sokendai), \\
National Astronomical Observatory of Japan, 2-21-1 Osawa, 
Mitaka, Tokyo, 181-8588}
\email{kaorin@solar.mtk.nao.ac.jp}
\altaffiltext{2}{National Astronomical Observatory of Japan, 
2-21-1 Osawa, Mitaka, Tokyo, 181-8588}
\altaffiltext{3}{W.W. Hansen Experimental Physics Laboratory, 
Stanford University, Stanford, CA 94305, USA}
\altaffiltext{4}{Department of Astronomy, School of Science,
University of Tokyo, Bunkyo-ku, Tokyo 113-0033}
\altaffiltext{5}{High Altitude Observatory, 
National Center for Atmospheric Research, P.O. Box 3000,
Boulder, CO 80307, USA}
\altaffiltext{6}{Kwasan and Hida Observatories, 
Kyoto University, Kurabashira, Kamitakara-cho,
Takayama, Gifu, 506-1314}
\altaffiltext{7}{Institute of Space and Astronautical Science, 
Japan Aerospace Exploration Agency,\\
3-1-1 Yoshinodai, Sagamihara, Kanagawa 229-8510}
\altaffiltext{8}{Lockheed Martin Solar and Astrophysics 
Laboratory, B/252, 3251 Hanover St., Palo Alto, CA 94304, U.S.A.}


\maketitle

\begin{abstract}
Exploiting high-resolution observations made 
by the Solar Optical Telescope 
onboard Hinode,
we investigate the spatial distribution of power spectral density
of oscillatory signal
in and around active region NOAA 10935.
The G-band data show that 
in the umbra the oscillatory power is suppressed in all frequency 
ranges. On the other hand, 
in Ca \emissiontype{II} H intensity maps
oscillations in the umbra, so-called umbral flashes,
are clearly seen with
the power peaking around 5.5 mHz.
The Ca \emissiontype{II} H power distribution 
shows the enhanced elements 
with the spatial scale of the
umbral flashes over most of the umbra
but there is a region with suppressed power 
at the center of the umbra.
The origin and property of this node-like feature 
remain unexplained.
\end{abstract}

\section{Introduction}

Oscillations within sunspots have been studied
since \citet{1969SoPh....7..351B} discovered the
umbral flashes which 
are transient brightenings in chromospheric lines in the umbra. 
In addition to oscillations in intensity
in various spectral lines,
fluctuations in velocity and magnetic field strength
were also observed.
So far, 
three types of oscillations have been observed within sunspots:
3 mHz (five-minute) oscillations in the photospheric umbrae,
5 mHz (and hight-frequency) oscillations in the 
chromospheric umbrae, and 
running penumbral waves (see review articles by
\cite{1985AuJPh..38..811T};
\cite{1992sto..work..261L}; \cite{1999ASPC..184..113S}; 
\cite{2000SoPh..192..373B} and references therein).
This classification may not be so distinct, as
some say that the chromospheric oscillations in umbrae 
(umbral flashes) 
and the running penumbral waves might be different manifestations
of the same phenomenon 
(\cite{2003A&A...403..277R}; \cite{2007A&A...463.1153T}).
Three-minute oscillations in the photosphere of umbrae
have also been reported, firstly by \citet{1972SoPh...27...61B}
although for their detection
a possibility of chromospheric contamination
has been pointed out since then (see \cite{1992sto..work..261L}).

How these oscillations are driven remains unestablished, 
however.
\citet{1992sto..work..261L} summarized in his review that
two theoretical pictures of 
the umbral oscillations in the chromosphere were drawn: 
one was that 
umbral oscillations are the resonant response of the atmosphere
to forcing by a broad-band power sources below,
and the other was that the oscillations are driven by a 
sub-photospheric resonance to fast-mode waves.
More recently,
\citet{2003A&A...403..277R} concluded that
the umbral flashes are near-acoustic field-guided 
upward-propagating shock waves.
\citet{2006ApJ...640.1153C} 
also found evidence for shock waves propagating 
upward in umbral regions.
Umbral oscillations were also observed in higher atmosphere;
\citet{2001ApJ...550.1113S} reported a radio brightness oscillation
of 3-minute period above a sunspot umbra and interpreted it as an
upward-traveling acoustic wave.
\citet{2007A&A...461L...1V} reported that
magnetic network elements 
can also channel low-frequency photospheric oscillations into 
chromosphere even in the quiet region \citep{2006ApJ...648L.151J}.

The issue of how the oscillations in sunspots 
are driven, and how the waves that are associated
with these oscillations propagate in magnetized atmosphere is 
important in probing subsurface structure of sunspots using
detailed observations of sunspot oscillations, which was 
first proposed by \citet{1982Natur.297..485T} and
now is an actively pursued goal of 
local helioseismology (see, e.g.,
\cite{2000SoPh..192..159K}).
Using local helioseismological technique,
one can extract information about subsurface 
local structures by measuring travel times for given distances of 
(primarily) acoustic waves.
We need detailed knowledge of how 
these sunspot waves are generated, and 
how they interact with magnetic field,
in our attempt to reveal how the 
active regions are generated, evolve, and dissipate
by local helioseismology.

Solar Optical Telescope (SOT; \cite{2007SoPh.SOT}) 
onboard Hinode \citep{2007SoPh.Hinode} reveals 
many fine structures in sunspots, such as 
the penumbral flows and the light bridges.
In this paper, we report initial SOT observations of 
sunspot oscillations in Ca \emissiontype{II} H line and G band
with unprecedented high resolution.
We observe the lower chromospheric oscillations 
using the Ca \emissiontype{II} H line data,
while the photospheric oscillations are investigated 
in the G band. 
As for helioseismic observation in the quiet region 
with SOT, see \citet{2007PASJ.Sekii}.
In section \ref{sec:obs}, we 
briefly describe the SOT observations
and the data reduction procedure, and what we found
on the umbral oscillations are described in 
section \ref{sec:result}.
Discussions mainly on the umbral oscillations are given 
in section \ref{sec:discussions}.

\section{Observations and Data Reduction}
\label{sec:obs}
\begin{figure}[hptb]
\begin{center}
\begin{tabular}{c}
\FigureFile(80mm,40mm){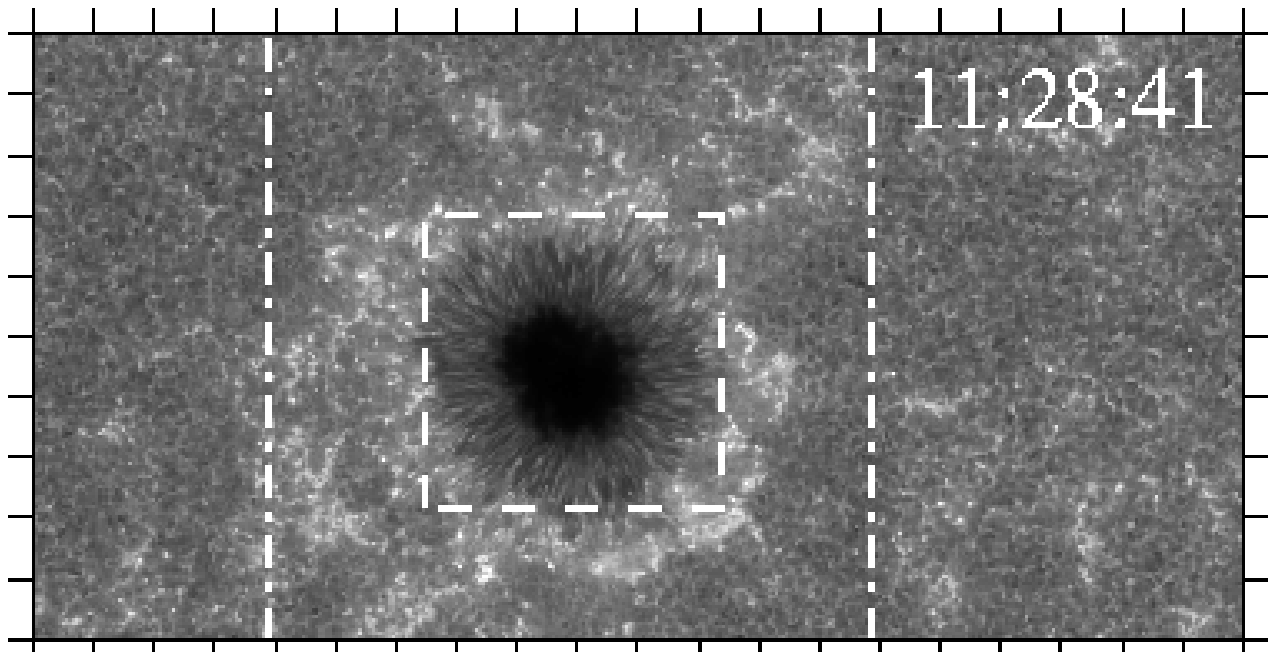}\\
\FigureFile(80mm,40mm){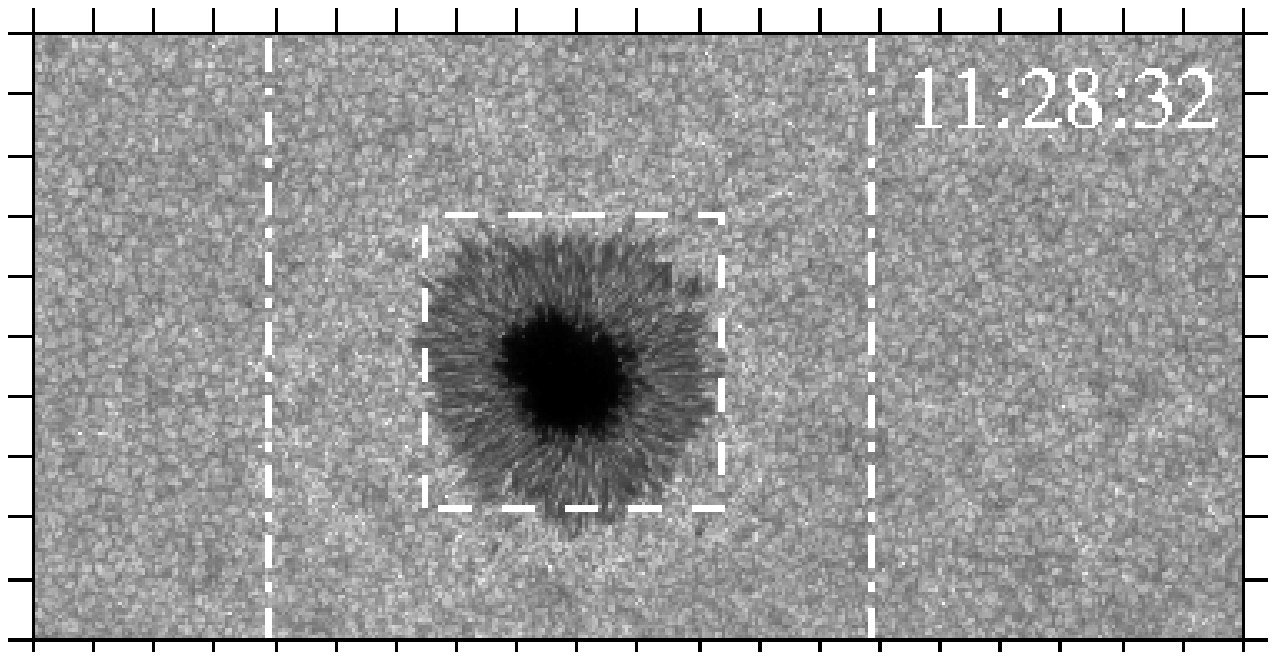} \\
\end{tabular}
\end{center}
\caption{Sample intensity maps in the Ca \emissiontype{II} H line (top)
and the G band (bottom).
We examined the intensity oscillations in 
this field of view (201 arcsec $\times$ 101 arcsec).
The ticks are spaced by 10 arcsec.
The central boxes with dashed lines indicate the 
field of view in figure \ref{fig:CaUmbra},
while the dot-dashed boxes indicate the 
field of view of the power maps in figures \ref{fig:pm_Ca}
and \ref{fig:pm_G}.}
\label{fig:sampleI}
\end{figure}

A fairly round sunspot close to the disk center (NOAA 10935) 
was observed with the Solar Optical Telescope (SOT)
onboard Hinode over a duration of 4 hr 42 minutes 
(11:18 - 16:00 UT) 
on January 8, 2007.
Using Broadband Filter Imager (BFI; \cite{2007SoPh.SOT}) of SOT, 
we obtained series of 218 arcsec $\times$ 
109 arcsec filtergrams 
in Ca \emissiontype{II} H (3968.5 \AA) and G band (4305 \AA)
with a cadence of $\sim$ 1 min.
The cadence was slightly irregular, although 
it hardly affects the Fourier analyses
we carry out (see \cite{2007PASJ.Sekii}). 
To reduce the data amount, 2 $\times$ 2 summing was 
carried out onboard;
hence, the pixel size was $\sim$ 0.1 arcsec. 
About 10 seconds after taking each G-band image, 
a Ca \emissiontype{II} H image was taken. 
During the period, 
the correlation tracker (CT; \cite{2007SoPh.CT}) 
was used to stabilize the images.
Since CT only estimates and compensates for 
the movement in the granules,
we need additional tracking to account for
the proper motion of the sunspot.
We used a two-dimensional cross-correlation technique
to compute the displacement. 
The displacement was then smoothed 
by polynomial fitting before applied to the series of images.
The difference between the original displacements 
and the smoothed ones were less than 0.7 arcsec,
and we consider this value as an estimate of the tracking error.
The field of view that we used for our analysis is 
slightly smaller (201 arcsec $\times$ 101 arcsec)
than the original one owing to this sunspot tracking.
Figure \ref{fig:sampleI} shows intensity maps in 
the Ca \emissiontype{II} H line and in the G band.
We studied the intensity oscillations in this field of view.

For dark subtraction, 
flat fielding, and correcting for bad pixels, 
we used a calibration program 
provided by the SOT team.  
Running difference images were used to
remove any possible remnant temporal or spatial trend.
Then, in each pixel, 
the intensity difference was normalized by the mean intensity 
in the two running frames.
This was mainly in attempt to detect oscillations 
even in the region with very low intensity, i.e., the umbra.
It also permits us to observe 
a smooth transition between umbra and penumbra.
We then used Fourier transform of the relative 
intensity difference time series at each pixel
to produce power-spectral-density maps.
The Nyquist frequency is 8.3 mHz in our analyses in
accordance with the 1-min cadence. 
Since taking the running difference is equivalent to taking
the time derivative on the discrete time grid, 
we divided the obtained power by squared 
angular frequency $\omega^2$.
Thus the main remaining effect of taking running difference
is the reduced spatial trend.

It should be noted that the power
maps measure the intensity variation due to both 
(magneto)convection
and oscillation. These cannot be distinguished by 
this kind of analysis,
except that the convection spectrum is expected to be 
continuous with a monotonic decrease with frequency.

Another technical point is that because the oscillation
spectra are broad-band, normalization by temporally local mean 
does not affect them very much.
We confirmed this by carrying out the analysis without the 
normalization, where we found no qualitative change,
aside from much reduced power in the umbra, which was to be 
expected.

\section{Results}
\label{sec:result}

\subsection{Umbral Oscillations observed in Ca \emissiontype{II} H}

\begin{figure}[h]
\begin{center}
\FigureFile(75mm, 58mm){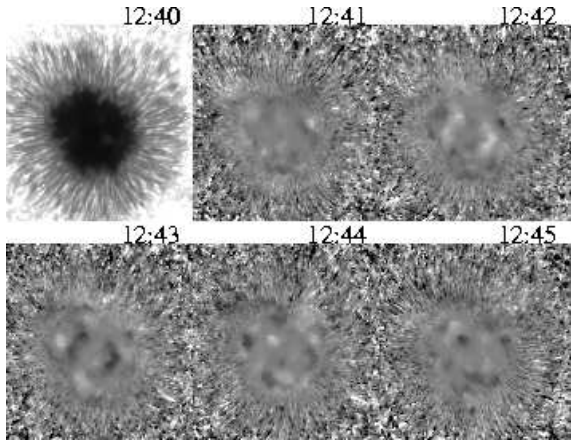}
\end{center}
\caption{A Ca \emissiontype{II} H intensity map (upper-left) and 
running difference intensity maps (the others)
showing umbral oscillations. 
White marks the region where 
the intensity increases compared with 
that in the previous frame.
The field of view is a 49-arcsec square.
The figure in the upper-right corner of each panel indicates time.
An extended version of these figures are 
available as an mpeg animation 
in the electronic edition.}
\label{fig:CaUmbra}
\end{figure}

Figure \ref{fig:CaUmbra} shows 
one example of umbral oscillations
in our data set.
Running difference intensity maps in Ca \emissiontype{II} H 
are shown in this figure, except the upper-left panel 
of the intensity map for reference.
Please note that, although here we present the running differences for 
visual enhancement, power spectra are always corrected for the 
$\omega^2$ factor.
Localized brightening randomly occurs in most of 
the umbra;
in a snapshot of the filtergram, small bright 
patches are seen.
These are ``umbral flashes'' 
(e.g., \cite{2007A&A...463.1153T}).
In the running difference intensity maps,
there are some elements with the size of 5 arcsec
where a bright region is adjacent to a dark region.
Since this means the region with enhanced intensity
has just moved from the dark region to the adjacent bright region,
we can roughly estimate the propagation 
speed of the flashes
based on separation between these structures;
the features moved $\sim$ 4 arcsec in a minute, so 
the propagation speed is up to 50 ${\rm km \ s^{-1}}$.
When we see the running difference movie,
these wave-like 
structure seem to move outward rather than inward.
Unlike in the Ca \emissiontype{II} H data,
we cannot find any umbral oscillations 
in the G-band intensity data.

\subsection{Relative Intensity Power Maps}
\begin{figure}[htpb]
\begin{center}
\FigureFile(80mm,160mm){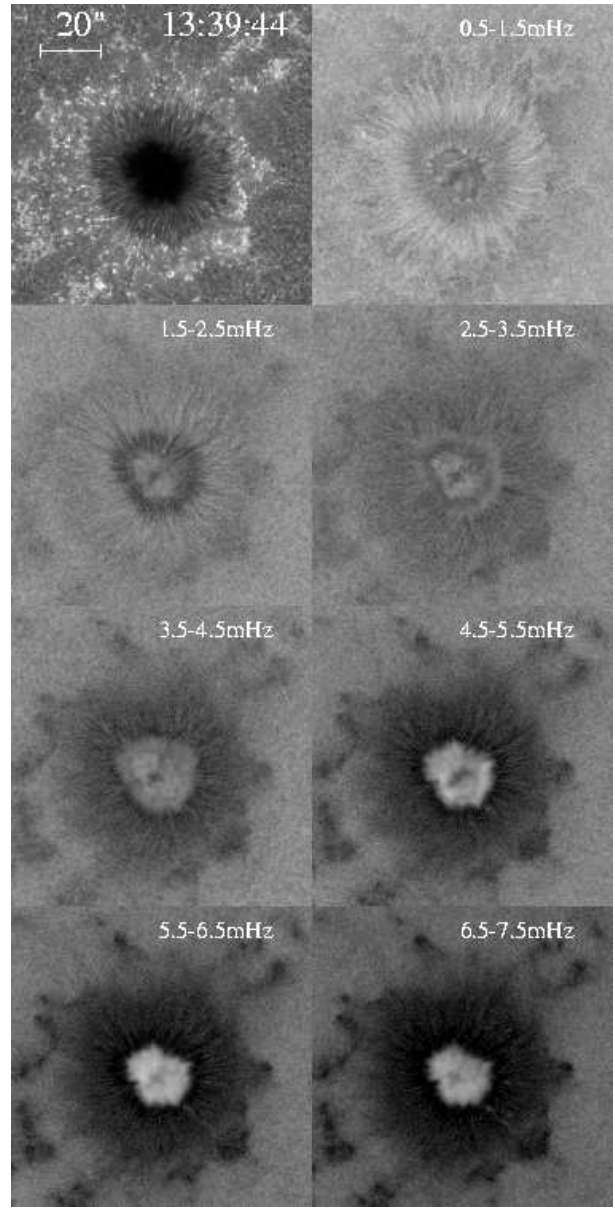}
\end{center}
\caption{A Ca \emissiontype{II} H intensity image 
(top-left) and the power maps 
from Ca \emissiontype{II} H intensity data
of active region NOAA 10935.
The field of view is 100 arcsec square in all the panels.
The power is displayed in logarithmic greyscaling and the 
same color range is used in
figures \ref{fig:pm_Ca}, \ref{fig:pm_average_Ca},
\ref{fig:pm_G}, and \ref{fig:pm_average_G}.}
\label{fig:pm_Ca}
\end{figure}

\begin{figure}[hbpt]
\begin{center}
\FigureFile(80mm,40mm){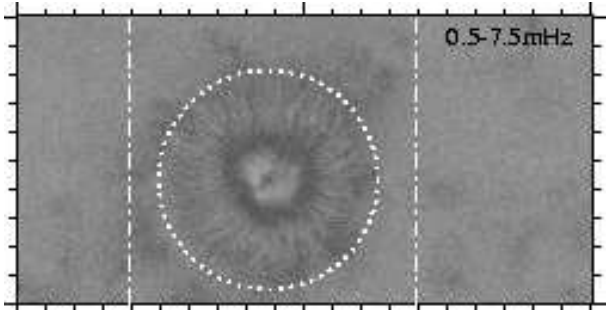}
\end{center}
\caption{Ca \emissiontype{II} H line 
relative intensity power maps averaged 
over the frequency range from 0.5 mHz to 7.5 mHz
in the full field of view.
The dot-dashed lines indicate the field of view in figure
\ref{fig:pm_Ca}.
The center of the dotted circle is located at the
center of gravity of the node-like dark region.
The radius of the circle is 38 arcsec, and 
the cross sections in the circle is shown in 
figure \ref{fig:pm_average_Ca_cross}.}
\label{fig:pm_average_Ca}
\end{figure}

\begin{figure}[hptb]
\begin{center}
\FigureFile(80mm,50mm){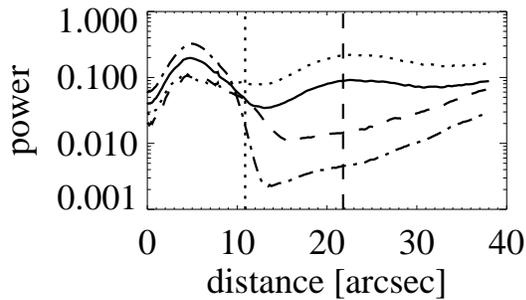}
\end{center}
\caption{Cross sections of the Ca \emissiontype{II} H power map
in the sunspot shown in 
figures \ref{fig:pm_Ca} and \ref{fig:pm_average_Ca}.
Distance is measured from the center of the node.
The solid line indicate the cross section of the 
power map averaged over the frequency range 
from 0.5 mHz to 7.5 mHz.
The dotted, dashed, and dot-dashed lines indicate the 
cross section of the power maps averaged over 
the 0.5--3.5 mHz, 3.5--4.5 mHz, and 4.5--7.5 mHz ranges,
respectively.
The umbra/penumbra boundary is marked by the dotted vertical line, 
and the dashed vertical line indicates the
penumbra/moat boundary.}
\label{fig:pm_average_Ca_cross}
\end{figure}

\begin{figure}[htpb]
\begin{center}
\FigureFile(80mm,160mm){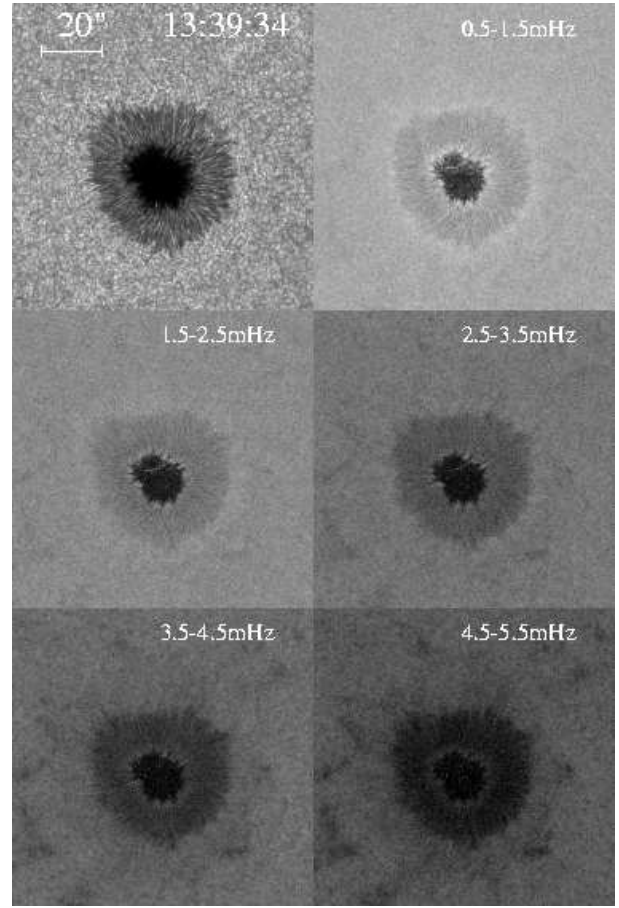}
\end{center}
\caption{A G-band intensity image 
(top-left) and the power maps 
from G-band intensity data of active region NOAA 10935.
The field of view is the same as that in figure \ref{fig:pm_Ca}.}
\label{fig:pm_G}
\end{figure}

\begin{figure}[hptb]
\begin{center}
\FigureFile(80mm,40mm){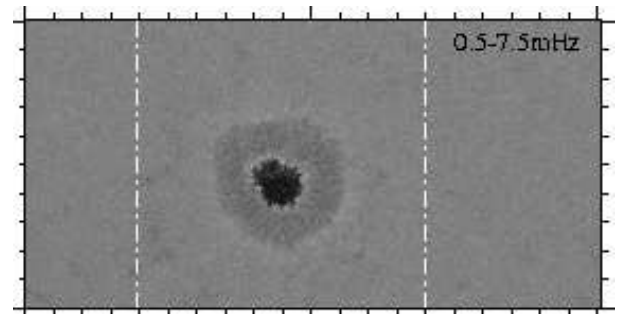}
\end{center}
\caption{G-band relative intensity power maps averaged 
over the frequency range from 0.5 mHz to 7.5 mHz
in the full field of view.
The dot-dashed lines indicate the field of view in
figure \ref{fig:pm_G}.}
\label{fig:pm_average_G}
\end{figure}

Since oscillations were seen in the Ca \emissiontype{II} H
running difference intensity movie,
we investigated these oscillations in detail by 
decomposing them into Fourier components.
Figures \ref{fig:pm_Ca} and \ref{fig:pm_average_Ca} show a
sample Ca \emissiontype{II} H intensity map and
the power maps derived from the Ca \emissiontype{II} H intensity data.
In figure \ref{fig:pm_Ca}, 
we show the relative intensity power maps averaged over 
1 mHz wide frequency ranges,
from 1 mHz to 7 mHz
with logarithmic greyscaling,
while in the figure \ref{fig:pm_average_Ca}
the power map averaged over the wider frequency range 
from 0.5 mHz to 7.5 mHz.
Figure \ref{fig:pm_average_Ca_cross} shows 
the cross sections of the power maps:
we consider circles around the center of the `node' 
(the center of the dotted 
circle shown in figure \ref{fig:pm_average_Ca})
and average the power on each circle to 
obtain the azimuthally averaged power distribution 
as a function of the distance from the center of the node.
The solid curve indicates the cross section of the 
averaged power map (figure \ref{fig:pm_average_Ca}),
while the dotted, dashed, and dot-dashed curves
show the cross section
of the power maps averaged over 
0.5--3.5 mHz, 3.5--4.5 mHz, 4.5--7.5 mHz ranges,
respectively.
Figures \ref{fig:pm_G} and \ref{fig:pm_average_G}
are the G-band counterparts of figures \ref{fig:pm_Ca}
and \ref{fig:pm_average_Ca}.
In figure \ref{fig:pm_G}, we omit the 
6 mHz and 7 mHz power maps, because
they are essentially identical to the 5 mHz power map, 
except throughout the maps
the power was smaller 
than that in 5-mHz range.
We find the following from inspecting these power maps:

1. In the Ca \emissiontype{II} H power maps, 
in all the frequency ranges,
there is a small area ($\sim$ 6 arcsec in diameter)
near the center of the umbra where the power was suppressed.
This node-like structure is seen more clearly in the 
power maps averaged over a wider frequency range 
(figure \ref{fig:pm_average_Ca})
and in its cross section 
(figure \ref{fig:pm_average_Ca_cross}).
This type of `node' has not been reported so far,
except perhaps the `calmest umbral position' found by 
\citet{2007A&A...463.1153T} in Doppler power maps
(however, for discussion of umbral oscillation patterns,
see \citet{1975PASJ...27..259U} who compared their
calculation with \authorcite{1972SoPh...27...71G}'s
(\yearcite{1972SoPh...27...71G}) observation).
Possibly, stable high-resolution observation made by Hinode/SOT 
was required to find such a tiny node,
although it is also possible that
only a particular type of sunspots, e.g., round ones
with axisymmetric geometry, exhibit such node-like structure;
we need to observe various types of sunspots to 
investigate the possible geometrical effects.
We discuss this node in more detail later 
(section \ref{sec:discuss_uf}). 

2. Above 4 mHz in the Ca \emissiontype{II} H power maps, 
power in the umbra is remarkably large.
In the power maps averaged over narrower frequency range
(0.05 mHz wide, not shown), 
the region with high power in the umbra seems to be more patchy. 
This probably corresponds to the elements of umbral flashes
mentioned in the previous subsection.
We discuss relationships between this structure and 
the umbral flashes in section \ref{sec:discuss_uf}.

3. In the lower-frequency range (1 mHz), the power is enhanced
at the umbra/penumbra boundary in G-band power maps. 
A similar feature 
was also reported by \citet{2001soho...10..219H};
a bright ring around a sunspot was seen in 
their SOHO/MDI intensity power map in the 0--1 mHz range.
This bright ring is discussed in section \ref{sec:discuss_br}.
In the higher-frequency ranges,
this bright region encircling the umbra
is also seen although less striking.
The dark regions in the power maps are 
smaller than the umbra in the intensity map
because of the bright ring.

4. In the Ca \emissiontype{II} H power maps,
a bright ring in the penumbra is found in 
lower frequencies, due to penumbral running waves.
\citet{1992sto..work..261L} reported that
chromospheric oscillatory frequency in the penumbra
decreases from 4 mHz at the inner boundary 
to less than 1.5 mHz at the outer boundary.
In the current data, this is seen
as the bright ring decreasing in size as the frequency increases.

5. The region with enhanced power in high-frequency ranges
around the active region, so-called ``acoustic halo''
(\cite{1992ApJ...392..739B}; \cite{1992ApJ...394L..65B}),
is not found in our power maps.
However, this is not inconsistent with the previous works,
because acoustic halos were reported to be
in the velocity power maps
obtained by Dopplergrams, but not in the
intensity power maps \citep{1998ApJ...504.1029H}.

6. In the regions that are outside the sunspot but are bright
in the Ca \emissiontype{II} H intensity
(such as moat region and plage-like features),
not only Ca \emissiontype{II} H power map
but also the G-band  power map, exhibits remarkable suppression
of signal in the higher frequency range, say above 5mHz.
Since these regions are
where the magnetic field strength is relatively strong, 
these features suggest that, outside the sunspot,
the power is suppressed as the magnetic field strength increases;
this confirms the previous view
(e.g., \cite{1998ApJ...504.1029H}; \cite{2002A&A...387.1092J}).

\subsection{Power Spectra}

\begin{figure}[htpb]
\begin{center}
\FigureFile(80mm, 70mm){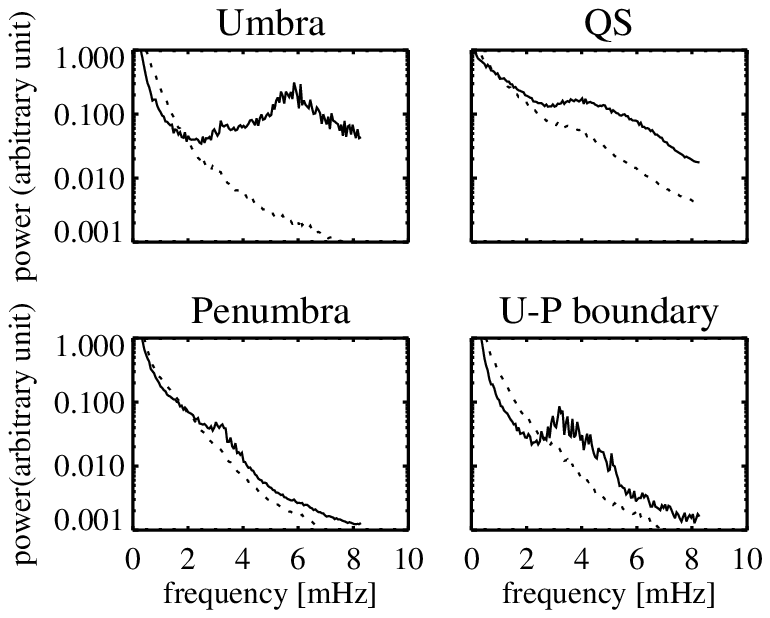}
\end{center}
\caption{Power spectra averaged in the umbra (upper left),
in the quiet region (upper right), 
in the penumbra (lower left), 
and around the boundary between the umbra and the penumbra
(lower right).
The Ca \emissiontype{II} H (solid) and
the G-band (dotted) intensity power spectra are shown.
The ordinate is in an arbitrary unit
in logarithmic scale.}
\label{fig:spectrum}
\end{figure}

To see how the  
power distributions in the umbra, penumbra, and the quiet 
region differ from each other, 
we examined the power spectrum in each region.
Figure \ref{fig:spectrum} shows 
the power spectra of  
the G-band and the Ca \emissiontype{II} H intensity oscillations. 
Each power spectrum is averaged over 39303 pixels 
(in the umbra), 44000 pixels (in the quiet region), 
62000 pixels (in the penumbra), or
21500 pixels (around the boundary between the umbra and penumbra).
Here, the umbra and penumbra were defined 
according to the average
Ca \emissiontype{II} H intensity over the observing period.
We defined the quiet region as the region that is outside the
sunspot and is not bright in Ca \emissiontype{II} H, and selected
a part of this region so that the area is comparable to those
of the umbra and penumbra.
The boundary between the umbra and penumbra
was defined as the annulus with the inner radius of 11 arcsec and
the outer radius of 14 arcsec. 
The center of the annulus is placed at
the center of the node (the center of the circle shown
in figure \ref{fig:pm_average_Ca}).

In all the regions, the G-band intensity power decreases almost
monotonically with the frequency,
except the broad peak around 4 mHz in the quiet region.
This peak corresponds to the global five-minute oscillation.
The reason for the lack of power excess 
in the umral G-band intensity is not understood.
The Ca \emissiontype{II} H intensity power in the 
quiet region shows the similar trend, 
while the power spectrum in the penumbra 
exhibits monotonic decrease, as is expected from 
the power maps in figure \ref{fig:pm_Ca}, 
except the narrow peak at 3 mHz.
The Ca \emissiontype{II} H intensity power spectrum in 
the umbra has two peaks: 
one around 3 mHz and the other around 5.5 mHz.
In the previous works
(e.g, a review by \cite{1992sto..work..261L}), 
the dominant period of oscillation in the chromosphere 
was above 5.5 mHz,
and, in contrast to our results, 
no significant power peaks were found in the 3 mHz range.
Comparison with the power spectrum around the boundary 
between the umbra and penumbra in figure \ref{fig:spectrum}
indicates that 
the 3 mHz peaks in the umbral and penumbral spectra
originate in the boundary region. 
The broad peak in the boundary region,
between 2 mHz and 5 mHz, is brought about by 
running penumbral waves.

\section{Discussions}
\label{sec:discussions}

\subsection{Bright Ring in the Power Maps}
\label{sec:discuss_br}

As was shown in the G-band 
power maps in figure \ref{fig:pm_G},
there was a bright ring-shaped structure around the
penumbra,
which appeared in all the frequency ranges,
and was most visible in the 1 mHz frequency range. 
A similar bright ring was reported by 
\citet{2001soho...10..219H}. 
Their power maps derived from SOHO/MDI intensity data
showed a bright ring around the sunspot in the 0--1 mHz range.
In their case,
a strong enhancement of power within a sunspot was also observed 
in Ca \emissiontype{II} K intensity
in all frequency ranges obtained by TON data
only to be identified as artifacts due to terrestrial seeing. 
In the case of space-borne observation, 
seeing never comes into play. 
There is some possibility that 
artificial oscillation signals are produced by a poor
tracking of the sunspot,
since at the sunspot boundary the intensity has large contrast.
However, at 0.7 arcsec, the tracking error is smaller than
the width of the bright ring around the penumbra ($\sim 3$ arcsec).
We therefore conclude the structures are not due to 
the tracking error.
The true nature of this ring is not known,
though it seems to be associated with various motions around
the umbra/penumbra boundary including penumbral bright filaments
breaking into the umbra, and boundary itself moving around by
up to $\sim 2$ arcsec, which is similar to the thickness of the
ring, during the observation. These motions themselves may be
manifestations of MHD oscillatory phenomena.

\subsection{Umbral Flashes}
\label{sec:discuss_uf}

Umbral flashes are localized transient brightenings 
in chromospheric umbrae.
Since \citet{1969SoPh....7..351B} discovered them 
in Ca \emissiontype{II} H and K lines,
many observations and theoretical works have been carried out
on the subject
(e.g.,  \cite{2003A&A...403..277R}; \cite{2007A&A...463.1153T}).
According to them, 
the period of the umbral flashes
is around 3 minutes,
and the brightening elements 
which appear from place to place in the umbrae
are 3--5 arcsec wide.
In our Hinode/SOT Ca \emissiontype{II} H data, 
we confirmed these properties, 
or at least found the correspondent features: 
the strong peak around 5.5 mHz in the umbral power spectrum,
and the patchy 
structure of the transient brightenings of about 5 arcsec
in the running difference intensity movie.  
It was also reported that they 
expand outward in the shape of arcs 
at the speed of 5--20 ${\rm km \ s^{-1}}$
and run into the penumbra.
As mentioned in section \ref{sec:result},
we find that they seem to move outward at a 
speed of up to 50 ${\rm km \ s^{-1}}$,
as a very rough estimate. For a better estimate,
a higher cadence is required.

So far the umbral flashes are interpreted 
as some kinds of magnetohydrodynamic (MHD) waves
\citep{1970SoPh...13..323H}
and upward-propagating shocks
(\cite{2003A&A...403..277R}; \cite{2006ApJ...640.1153C}) .
Since umbral flashes are known to occur only 
when the velocity amplitudes exceed a certain threshold
(see, e.g., \cite{2007A&A...463.1153T}),
\citet{2003A&A...403..277R} suggested that a
large Dopplershift due to 
upward propagation along the line-of-sight
results in a drop of opacity at the line center and
causes the brightness enhancements, i.e., the umbral flashes. 
However, these models do not readily explain the 
node-like structure we find around the center of the umbra
as one then expects the umbral flashes 
are the most visible in the umbral center.
Similar feature to our node-like structure was reported by
\citet{2007A&A...463.1153T}.
In their Doppler velocity maps in Ca \emissiontype{II} 8542 \AA \
there was a small area around the center of the umbra
where the velocity amplitude was lowest. In their paper,
intensity fluctuation in
the `calmest umbral position' was suggested to be 
associated with velocity perturbation
spreading from the surrounding umbral
flashes by means of running waves
rather than upward-propagating shocks under conditions
different from those in the umbral flashes.

It has to be mentioned that 
the running umbral waves in the chromosphere reported by 
\citet{2004A&A...424..671K} may be the same oscillational
phenomenon as the umbral flashes. 
In their study, using H$\alpha$ data instead of 
Ca \emissiontype{II} data, they found line-of-sight 
velocity oscillations in the sunspot umbra,
but they did not find the brightenings in H$\alpha$.
Therefore, they concluded that the running umbral waves they found
was different from the umbral flashes. However, 
as they mentioned, brightening in H$\alpha$ does not
always accompany the umbral flashes; 
since the other properties of the running umbral waves 
are similar to those of 
the umbral flashes, we cannot exclude the possibility
that they are the same phenomenon.

Compared with oscillation signal detected in Dopplergrams,
oscillation signal detected in intensity maps 
is not easily interpretable,
because the intensity oscillations are affected not only by the 
motion of the plasma, but also by how the 
fluctuation of density, temperature,
degree of ionization, and
other thermodynamic quantities affect the line formation.
Detailed radiative transfer calculation needs to
be undertaken.
For further study,
we need to compare the oscillations in intensity maps
with those in Dopplergrams to understand the 
relationships between the oscillation parameters.


\bigskip
Hinode is a Japanese mission developed and launched 
by ISAS/JAXA, collaborating with NAOJ as a domestic partner, 
NASA and STFC (UK) as international partners. Scientific operation 
of the Hinode mission is conducted by the Hinode science team 
organized at ISAS/JAXA. This team mainly consists of scientists 
from institutes in the partner countries. 
Support for the post-launch 
operation is provided by JAXA and NAOJ (Japan), 
STFC (U.K.), NASA, ESA, and NSC (Norway).
This work was carried out at the NAOJ Hinode Science Center,
which is supported by the Grant-in-Aid for 
Creative Scientific Research 
``The Basic Study of Space Weather Prediction'' from MEXT, Japan 
(Head Investigator: K. Shibata), generous donations 
from Sun Microsystems, and NAOJ internal funding.
K.~Nagashima is supported by the Research Fellowship from the 
Japan Society for the Promotion of Science for Young Scientists.



\end{document}